\documentclass[fleqn,usenatbib]{mnras}

\usepackage{newtxtext,newtxmath}

\usepackage[T1]{fontenc}
\usepackage{hhline}
\DeclareRobustCommand{\VAN}[3]{#2}
\let\VANthebibliography\thebibliography
\def\thebibliography{\DeclareRobustCommand{\VAN}[3]{##3}\VANthebibliography}

\usepackage{graphicx}	
\usepackage{amsmath}	

\usepackage{amssymb}	

\usepackage[capitalise]{cleveref}
\usepackage{lipsum}
\newcommand{\pkg}[1]{\texttt{#1}}
\newcommand{\fastsim}{\pkg{21cmFAST}}

\newcommand{\lya}{Ly$\alpha$}

\title[HERA Cross-Correlation]{Estimating the Feasibility of 21cm-\lya\ Synergies using the Hydrogen Epoch of Reionization Array}
\author[Cox et. al]{
Tyler A. Cox,$^{1, 2}$\thanks{E-mail: tyler.a.cox@berkeley.edu}
Daniel C. Jacobs,$^{2}$
Steven G. Murray,$^{2}$
\\
$^{1}$Department of Astronomy, University of California, Berkeley, CA\\
$^{2}$School of Earth and Space Exploration, Arizona State University, Tempe, AZ
}
\date{Accepted 2022 February 15. Received 2022 February 14; in original form 2021 March 30}
\pubyear{2021}
\graphicspath{{./}{figures/}}

\begin{document}
\label{firstpage}
\maketitle

\begin{abstract}

Cross-correlating 21cm and Ly$\alpha$ intensity maps of the Epoch of Reionization (EoR) promises to be a powerful tool for exploring the properties of the first galaxies. Next-generation intensity mapping experiments such as the Hydrogen Epoch of Reionization Array (HERA) and SPHEREx will individually probe reionization through the power spectra of the 21cm and \lya \ lines respectively, but will be limited by bright foregrounds and instrumental systematics. Cross-correlating these measurements could reduce systematics, potentially tightening constraints on the inferred astrophysical parameters. In this study, we present forecasts of cross-correlation taking into account the effects of exact uv-sampling and foreground filtering to estimate the feasibility of HERAxSPHEREx making a detection of the 21cm-Ly$\alpha$ cross-power spectrum. We also project the sensitivity of a cross-power spectrum between HERA and the proposed next-generation Cosmic Dawn Intensity Mapper. By isolating the sources of uncertainty, we explore the impacts of experimental limitations such as foreground filtering and \lya\ thermal noise uncertainty have on making a detection of the cross-power spectrum. We then implement this strategy in a simulation of the cross-power spectrum and observational error to identify redshifts where fiducial \fastsim\ models predict the highest signal-to-noise detection ($z \sim 8$). We conclude that detection of the SPHEREx-HERA cross-correlation will require an optimistic level of 21cm foreground filtering, as well as deeper thermal noise integrations due to a lack of overlapping sensitive modes but for CDIM with its larger range of scales and lower noise forecast detection levels, may be possible even with stricter 21cm foreground filtering.

\end{abstract}


\section{Introduction} \label{sec:intro}
The Epoch of Reionization (EoR) marks a major phase transition in the history of the Universe. During this time period, the first stars and galaxies, formed from density fluctuations seeded by inflation, produced X-ray and ultraviolet photons, which heated and ionized the neutral gas around them. Gradually, these first luminous sources ionized all of the neutral hydrogen gas in the intergalactic medium (IGM) around them, transitioning the ionization state of the Universe from completely neutral to fully ionized (see \citealt{2001PhR...349..125B} for a review). 
Only a limited number of observations have been made which constrain reionization timing and the properties of the sources driving its progression. Observations of the Gunn-Peterson trough \citep{1965ApJ...142.1633G} in the spectra of high-redshift quasars have been used to place the conclusion of reionization at $z \sim 6$ \citep{2006ARA&A..44..415F}. Assuming a model in which reionization took place instantaneously, the integrated Thomson optical depth of cosmic microwave background photons has also been used to imply a reionization redshift of $z \sim 7.68$ (\citealt{2020A&A...641A...5P, 2020A&A...641A...6P}).  These constraints suggest that the bulk of reionization took place from $z \sim 6 - 10$, yet little is known about the topology of reionization, the properties of the first structures, or the spatial or temporal evolution of the ionized bubbles surrounding the first luminous sources.

New techniques are currently being developed to help shed light on this period of time. One of the most promising techniques for directly observing the EoR is intensity mapping of the 21cm hyperfine transition of neutral hydrogen  (for a review of 21cm cosmology, see \citealt{2006PhR...433..181F}). The potential for understanding reionization through the 21cm line has spawned a number of experiments over the past decade, including PAPER\footnote{Precision Array for Probing the Epoch of Reionization \citep{2019ApJ...883..133K}}, LOFAR\footnote{Low Frequency Array\citep{2019MNRAS.488.4271G}}, the GMRT\footnote{Giant Metre Wave Radio Telescope \citep{2013MNRAS.433..639P}}, and MWA\footnote{Murchison Widefield Array \citep{2019ApJ...887..141L}}, which have set increasingly stringent upper limits on the amplitude of the 21cm power spectrum. Building on techniques developed in previous experiments, the Hydrogen Epoch of Reionization Array (HERA; \citealt{2017PASP..129d5001D}) ---currently being built in South Africa---is expected to detect and characterize the 21cm power spectrum with high significance within the coming decade. However, while improvements have been made in collecting area and observing technique, HERA is still expected to be limited by the ability of the instrument to separate foregrounds and background at a level of 1 part in $\sim 10^5$.

Complementary to the 21cm line is intensity mapping of highly-redshift Lyman-$\alpha$ (\lya) emission which directly measures high-redshift galaxies and, indirectly, their effect on the ionized IGM \citep{2013ApJ...763..132S}. Previous work has suggested that wide-field intensity mapping of the \lya \ line may soon be possible with the future infrared satellites SPHEREx \citep{2014arXiv1412.4872D} and Cosmic Dawn Intensity Mapper (CDIM; \citealt{2016arXiv160205178C}).

Cosmological 21cm emission from the IGM is expected to be anti-correlated with galactic \lya\ at large spatial scales enabling study of the interface between ionized regions dominated by \lya\ and neutral regions emitting 21cm. Additionally, due to the fact that radio and infrared foregrounds are largely uncorrelated, cross-correlating these two measurements may help reduce problems associated with foreground removal, potentially leading to higher significance detections of the EoR.

Early studies of cross-correlation between 21cm and \lya\ emitters (LAEs) suggest at its potential for understanding the formation of the first galaxies \citep{2007ApJ...660.1030F, 2009ApJ...690..252L}. 
More recent work has shown the feasibility of cross-correlation between the Square Kilometer Array (SKA) and the Subaru Hyper Supreme Cam (\citealt{2017ApJ...836..176H, 2018MNRAS.479.2754K, 2018MNRAS.479.2767Y, 2020MNRAS.494..703W, 2020MNRAS.494.3131K}), and SKA and CDIM (\citealt{2017ApJ...846...21F, 2017ApJ...848...52H}). 
In particular, these latter papers demonstrated through both analytical (\citealt{2017ApJ...846...21F}) and semi-numerical modeling (\citealt{2017ApJ...848...52H}) that the cross-power spectrum is sensitive to astrophysical parameters associated with reionization, such as the mean free path of ionizing photons, and can potentially be used to set tighter constraints on those parameters than 21cm or \lya\ observations could provide independently. Similarly, \citealt{2016MNRAS.459.2741S} showed that cross-correlation between 21cm and LAEs improves constraints on the inferred volume-averaged neutral fraction.

While these results are highly encouraging, many of these estimates use instruments that may be a decade or more away from construction, delaying a detection of the 21cm-\lya\ until the mid-2030's. In this paper, we explore the feasibility of probing the EoR using the 21cm-\lya\ cross-power spectrum by combining HERA observations with the next-generation infrared probes SPHEREx and CDIM. HERA and SPHEREx are expected to deliver observations within the next decade. Given their overlapping survey area and sensitivity of each of these experiments, HERA and SPHEREx may offer the first opportunity to cross-correlate reionization-era intensity mapping measurements, which should help tighten constraints on the neutral fraction and independently confirm each detection. While a cross-power spectrum measurement made by HERA and SPHEREx should provide some constraints on the neutral fraction, the total signal-to-noise will likely be low. In order to fully explore HERA's potential to cross-correlate with other infrared instruments, we also provide projections for cross-correlation between HERA and CDIM. To make this estimate, we use the wide-field sensitivity for CDIM and imagine a scenario in which a CDIM wide-field survey overlaps with HERA's field of view.

This paper is organized as follows. In \textsection \ref{sec:modeling}, we discuss the model used for simulating intensities of line fluctuations, including a model for attenuation of Ly$\alpha$ by the neutral IGM. We then establish the notation and formally describe the cross-power spectrum and cross-correlation coefficient in \textsection \ref{sec:cross-power}. In \textsection \ref{sec:uncertainty}, we describe the uncertainty associated with a measurement of the cross-power spectrum, while including a treatment of the 21cm foregrounds, the exact layout of HERA, and thermal noise contributions from each experiment. Finally, we discuss the implications of our results and conclude in \textsection \ref{sec:summary}. Throughout this work, we assume a standard flat $\Lambda$CDM cosmology with the following parameters: $\Omega_{\Lambda} = 0.69$, $\Omega_{\rm m} = 0.31$, $\Omega_{\rm b} = 0.049$, $h = 0.68$, $n_s = 0.97$, and $\sigma_8 = 0.81$, which is consistent with the latest CMB measurements \citep{2020A&A...641A...6P}.

\section{Modeling Line Intensities} \label{sec:modeling}
Our goal is to establish the detectability and recoverable information content given a reasonable simulated prediction of 21cm and \lya{}. Significant progress has been made towards accurately modeling the conditions of the neutral IGM and \lya{} emitter population during reionization \citep{2006ApJ...653..815M, 2007MNRAS.377.1175D, 2010ascl.soft10025S, 2016MNRAS.463.2583K}.

The major challenge in modeling reionization is accurately capturing the physical processes that occur at vastly different spatial scales. While N-body/radiative transfer codes most accurately capture the physical properties and the evolution of reionization from relatively small scales in galaxies and large scales in the IGM, they are computationally expensive and are difficult to extend to the larger cosmological volumes that next-generation intensity mapping experiments will attempt to observe. Semi-numerical simulators are much more computationally efficient at modeling the evolution of the IGM and strongly agree with more numerically motivated simulators at the large scales probed by intensity mapping experiments within a few percent. Additionally, computational efficiency is not only more convenient, but necessary for the efficient exploration of parameter space required for robust parameter inference.

Here use the semi-numerical code, \fastsim \footnote{\url{https://github.com/21cmFAST/21cmFAST}} \citep{2011MNRAS.411..955M,Murray2020}, to simulate 21cm emission and to generate the halo catalogue, IGM density, and ionization fields necessary for the calculation of \lya\ emission in the following subsections.

    \subsection{21cm Brightness Temperature} \label{subsec:21cm}
\begin{figure*}
\centering
\includegraphics[width=.95\textwidth]{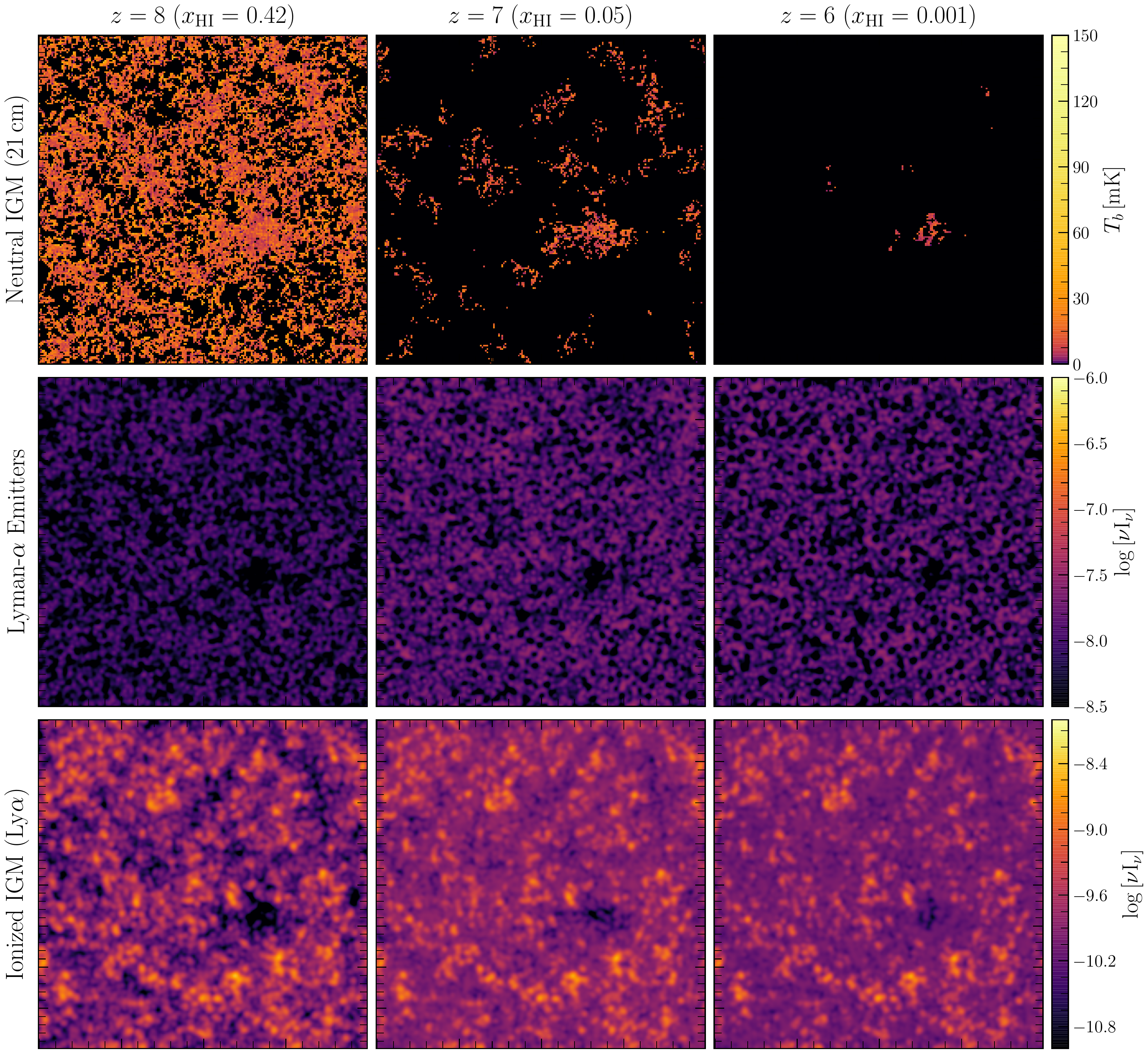}
\caption{Slices through the simulated 21cm brightness temperature (top), and Ly$\alpha$ emission from galaxies (middle) and the ionized IGM (bottom) at redshifts $z=8.06 \left(x_{\rm HI} = 0.42\right)$, $z=7.04 \left(x_{\rm HI} = 0.05\right)$, and $z=6.0 \left(x_{\rm HI} = 0.001 \right)$. The box size depicted is 300 Mpc in length with a resolution of 200 cells on each side. Further details of these simulations are described in \textsection \ref{subsec:21cm}, \ref{subsubsec:lyman-emitters}, and \ref{subsubsec:ion_IGM} respectively.\label{fig:sims}}
\end{figure*}
\fastsim\ is a semi-numerical simulator of the brightness temperature field as it evolves with cosmic time. It uses second-order Lagrangian perturbation theory (2LPT) to evolve a set of initial dark matter density perturbations to a given redshift. It then uses the excursion set formalism \citep{1991ApJ...379..440B, 2004ApJ...613....1F} to identify dark matter haloes in this field, and prescribes a hydrogen ``neutral fraction'' $x_{\rm HI}$ per-cell via a series of theoretically- and empirically-motivated relations. For more details of this procedure see \cite{2011MNRAS.411..955M}.

The 21cm brightness temperature is observed as a delta on the background CMB 
\begin{align}
  \delta T_b \left(z \right) & = \frac{T_s - T_{\gamma}}{1+z} \left(1 - e^{-\tau_{\nu_0}} \right ) \nonumber \\
      & \approx 27 x_{\rm HI} \left( 1 + \delta_{\rm nl} \right) \left( \frac{H}{dv_r/dv + H}\right) \left( 1 - \frac{T_{\gamma}}{T_s}\right) \nonumber \\
      & \hspace{1em} \times \left( \frac{1 + z}{10} \frac{0.15}{\Omega_{\rm M} h^2} \right)^{1/2} \left( \frac{\Omega_{\rm b} h^2}{0.023} \right) \textrm{ mK.}
      \label{eqn:offset_temp}
\end{align}
where $T_S$ is the gas spin temperature, $T_{\gamma}$ is the CMB temperature, $\tau_{\nu_0}$ is the optical depth at 21cm frequency, $\delta_{\rm nl} = \rho / \bar{\rho}_0 - 1$ is the non-linear density contrast, $H \left( z \right)$ is the Hubble parameter, $dv_r / dr$ is the comoving gradient of the line of sight component of the comoving velocity, where all quantities are evaluated at redshift $z = \nu_0 / \nu - 1$. The approximation in the second line of Eq. \ref{eqn:offset_temp} makes the assumption that reionization has reached the post-heating regime and that the CMB temperature, $T_{\gamma}$, is much smaller than the spin gas temperature, $T_S$. This approximation allows us to neglect the full spin gas temperature evolution through reionization when calculating $\delta T_b$, which is much more computationally efficient. Using the brightness temperature offset, we then calculate the 21cm fluctuation field, which will be used for cross-correlation in later sections, with the equation below,
\begin{equation}
\delta_{21} \left( \mathbf{x}, z\right) = \frac{ \delta T_b \left( \mathbf{x}, z\right)}{\overline{\delta T}_{b} \left( z \right)} - 1.
\end{equation}
Here $\overline{\delta T}_{b}$ is the spatial average of the 21cm brightness temperature.

For our simulations we used the fiducial model parameters utilized in \fastsim\ v3 with a box size of $\left(300 \rm Mpc\right)^3$. The simulated 21cm brightness temperature offset field defined in \cref{eqn:offset_temp} can be seen in the top row of Figure \ref{fig:sims}. \lya\ emission from galaxies (\textsection \ref{subsubsec:lyman-emitters}) and the ionized IGM (\textsection \ref{subsubsec:ion_IGM}) can be seen in the following two rows of the same figure for reference.

    \subsection{Near-Infrared Background} \label{subsec:lya}
Armed with a 21cm brightness temperature model we turn to generation of a self-consistent simulated \lya\ field which we do by post-processing the outputs of \fastsim.
We adopt the technique developed in \cite{2013ApJ...763..132S} and first applied to \fastsim\ simulations in \cite{2017ApJ...848...52H} for simulating \lya\ intensity mapping measurements. While those works already provide in-depth descriptions of the methods used to simulate \lya\ emission, we will restate them here for clarity. In this procedure, \lya\ fluctuations can be modeled from two distinct sources. Their origins are:

\begin{itemize}
    \item \textbf{Ly$\alpha$ Emitters}: emission from within the virial radius of dark matter halos. The dominant components are hydrogen recombinations and collisional excitation in galaxies.
    \item \textbf{Ionized IGM}: emission from the bubble of ionized gas that surround \lya-emitting galaxies. Here, recombinations of ionized hydrogen are the dominant contributor to \lya\ emission.
\end{itemize}

\subsubsection{Ly$\alpha$ Emitters} \label{subsubsec:lyman-emitters}

As mentioned above, the emission of \lya\ photons in LAE's is a result of two dominant processes: the recombination of ionized hydrogen and collisional excitations of neutral hydrogen within the virial radius of the halos. \cite{2013ApJ...763..132S} modeled two additional contributions to \lya\ emission but found them to be subdominant, so for this work we just focus on these two main sources. We predict both types of emission by post-processing the outputs from \fastsim{} which models galaxies and the neutral hydrogen IGM.

Both of these sources of emission are closely related to star formation and are therefore dependent on the star formation rate (SFR) of the LAEs which depends in turn on halo mass. We model SFR following \cite{2013ApJ...763..132S} who extrapolated down from the observed SFR of higher mass ( $M > 10^{11} M_{\odot}$) halos following the empirical model
\begin{equation}
  \mathrm{SFR} = A \!\left(\frac{M}{M_{\odot}} \right)^a \!\left( 1 + \frac{M}{c_1}\right)^{b} \!\left(1 + \frac{M}{c_2} \right)^d M_{\odot} \ {\rm yr}^{-1}
\end{equation}
where $A = 2.8 \times 10^{-28}$, $a = 2.8$, $b = -0.94$, $c_1 = 10^9 M_{\odot}$, $c_2 = 7 \times 10^{10} M_{\odot}$, and $d = -1.7$.

The dominant component to \lya\ emission in galaxies is the recombination of ionized hydrogen. As an electron cascades down energy levels during recombination, it has some probability of emitting a \lya\ photon. The number of \lya\ photons being emitted per second through these recombinations can be estimated using the relationship,
\begin{equation}
  \dot{N}_{\rm Ly\alpha} \left( M, z\right) = A_{\rm He} f_{\rm rec} f_{\rm Ly\alpha} \left[ 1 - f_{\rm esc} \left( M, z\right) \right] \dot{N}_{\rm ion}. \label{eqn:ndot}
\end{equation}
Here $A_{\rm He} = \left(1 - Y_{\rm He} \right) / \left( 1 - 3 Y_{\rm He}/4 \right)$ for a helium mass fraction $Y_{\rm He} = 0.249$, $f_{\rm rec}$ is the fraction of hydrogen recombinations that result in a \lya\ photon emission, $f_{\rm Ly\alpha}$ is the fraction of \lya\ photons not absorbed by dust, $f_{\rm esc}  \left(M, z\right)$ is the fraction of ionizing photons that escape the halo dependent on the halo mass following the relationship
\begin{equation}
  f_{\rm esc} \left( M, z\right) = \exp\left[-\alpha\left(z\right) M^{\beta\left(z\right)}\right]
\end{equation}
where $\alpha\left(z\right)$ and $\beta\left(z\right)$ are fitted values as a function of redshift \cref{table:fitted_values}, and $\dot{N}_{\rm ion} = Q_{\rm ion} \times \mathrm{SFR}$ is the rate of ionizing photons emitted by stars. The value for the average number of photons produced per solar mass of star formation, $Q_{\rm ion} \approx 5.8 \times 10^{60} \ M_{\odot}^{-1}$, was found by modeling population II stellar lifetimes and estimating the number of ionizing photons per unit time \citep{2002A&A...382...28S}.

We estimate a fraction $f_{\rm rec} \approx 66 \%$ of the hydrogen recombinations result in the emission of a \lya \ photon by making the assumption that clouds of interstellar gas are roughly spherical, that the gas temperature is of the $10^4 \rm K$, and an ionizing equilibrium in the gas is present as was calculated in \cite{1996ApJ...468..462G}. The \lya \ escape fraction, $f_{\rm Ly\alpha}$, is one that is more challenging to estimate as the value changes from galaxy to galaxy. For this study, we assume a redshift parameterization of the \lya \ escape fraction that was found in \cite{2011ApJ...730....8H},

\begin{equation}
    f_{\rm Ly\alpha} = C_{\rm dust} \times 10^{-3} \left(1 + z\right)^{\xi},
\end{equation}
where $C_{\rm dust} = 3.34$ and $\xi = 2.57$. This parameterization was devised such that $f_{\rm Ly\alpha}$ accounts for the difference between \lya \ luminosities found by scaling star formation rates and observed \lya \ luminosities assuming that the \lya \ photons emitted are a result of recombinations exclusively. It has been indicated in previous work that the $f_{\rm Ly\alpha}$ is not only a strong function of redshift, but is dependent on the halo mass as well, with $f_{\rm Ly\alpha}$ decreasing with increasing halo mass \citep{2011MNRAS.415.3666F}. As has been done in previous studies \citep{2013ApJ...763..132S, 2017ApJ...848...52H, 2017ApJ...846...21F}, we stick to the redshift dependent power law parameterization of the escape fraction as we do not expect it to the be dominant source of modeling error. We leave further parameterizations incorporating \lya \ escape fraction dependence on halo mass to future work.

\begin{table}
\centering
\begin{tabular}{ccccl}
\hhline{====}
\multicolumn{1}{l}{$z$} & \multicolumn{1}{l}{$\alpha\left(z\right)$} & \multicolumn{1}{l}{$\beta\left(z\right)$} & \multicolumn{1}{l}{$f_{\rm esc}\left(M = 10^{10} M_{\odot}, z\right)$} &  \\ \cline{1-4}
10.4                    & $2.78 \times 10^{-2}$        & 0.105                       & 0.73                                                                      &  \\
8.2                     & $1.3 \times 10^{-2}$         & 0.179                       & 0.45                                                                      &  \\
6.7                     & $5.18 \times 10^{-3}$        & 0.244                       & 0.24                                                                      &  \\
5.7                     & $3.42 \times 10^{-3}$        & 0.262                       & 0.24                                                                      &  \\ \cline{1-4}
\end{tabular}
 \caption{Escape fraction of UV radiation as a function of redshift fitted in \protect\cite{2010ApJ...710.1239R}. For simulation cubes whose redshift falls between these fitted redshift bins, we interpolate $\alpha\left(z\right)$ and $\beta\left(z\right)$ to estimate the UV escape fraction as a function of halo mass.}
\label{table:fitted_values}
\end{table}

To calculate the luminosity of \lya\ emission from galaxies, we multiply the rate of \lya\ photons being emitted by the energy of \lya\ photons to given us an expression for \lya\ recombination luminosity of a galaxy with a given mass, $M$, at redshift, $z$,
\begin{equation}
  L_{\mathrm{rec}}^{\mathrm{gal}} \left( M, z\right) = E_{\rm Ly\alpha} \dot{N}_{\rm Ly\alpha} \left( M, z \right),
\end{equation}
where we assume the emission of \lya\ radiation at rest-frame frequency with energy $E_{\rm Ly\alpha} = 13.6 \ {\rm eV}$.

The other dominant contributor of \lya\ emission in galaxies is excitation of neutral hydrogen. The \lya\ luminosity in the interstellar medium due to these excitations is defined as,
\begin{equation}
  L_{\mathrm{exc}}^{\mathrm{gal}} \left( M, z\right) = A_{\rm He} f_{\rm Ly\alpha} \left[ 1 - f_{\rm esc} \left( M, z\right) \right] E_{\rm exc} \dot{N}_{\rm ion} \left(M, z\right),
\end{equation}
where all terms have been previously defined, with the exception of $E_{\rm exc} \approx 2.14 \ {\rm eV}$, which was determined by estimating the average ionizing photon energy for thermal equilibrium, $E_{\nu} = 21.4$ eV \citep{2013ApJ...763..132S}, and relating it to the energy emitted as \lya\ radiation due to excitations, $E_{\rm exc} / E_{\nu} \approx 0.1$ \citep{1996ApJ...468..462G}.

With both dominant contributions to \lya\ emission modeled, we can then combine the contributions from excitation and recombination in galaxies to find a total \lya\ luminosity for galaxies dependent on halo mass and redshift,  $L^{\mathrm{gal}} = L_{\mathrm{exc}}^{\mathrm{gal}} + L_{\mathrm{rec}}^{\mathrm{gal}}$. Using this expression and the halo catalogue generated by \fastsim, we assign a \lya\ luminosity to each one of the halos in the catalogue. We then create a cube of \lya\ luminosity due to the contribution from galaxies that matches the voxel resolution of our $\delta T_b$ simulation cubes by adding the contribution of each halo to their corresponding voxel position, \textbf{x}. The result is a \lya\ luminosity cube whose emission distribution is naturally dependent on the spatial distribution of the halos and whose amplitude is influenced by the mass and clustering of the halos.

This luminosity cube is then converted to a luminosity density by dividing the galactic contribution to the \lya\ luminosity by the comoving voxel size of the simulation cube, $\ell^{\rm gal} = L^{\rm gal} / V_{\rm vox}$. This can be converted to a surface brightness using the expression
\begin{equation}
  I_{\nu}^{\rm gal}\left(\textbf{x}, z \right) = y \left(z \right)D_A^2 \left(z \right) \frac{\ell^{\rm gal} \left(\textbf{x}, z \right)}{4 \pi D_L^2}.\label{eqn:brightness}
\end{equation}
Here $D_A$ is the comoving angular diameter distance, $D_L$ is the luminosity distance, and the conversion factor from frequency to comoving distance is $y \left( z \right) = \lambda_{0} \left( 1 + z\right)^2 / H \left(z \right)$ (for the rest-frame wavelength of \lya\ radiation, $\lambda_0 = 1216$ \AA). Slices of these simulation cubes across the redshift range of interest can be found in Figure \ref{fig:sims}.

\begin{figure*}
\centering
\includegraphics[width=.95\textwidth]{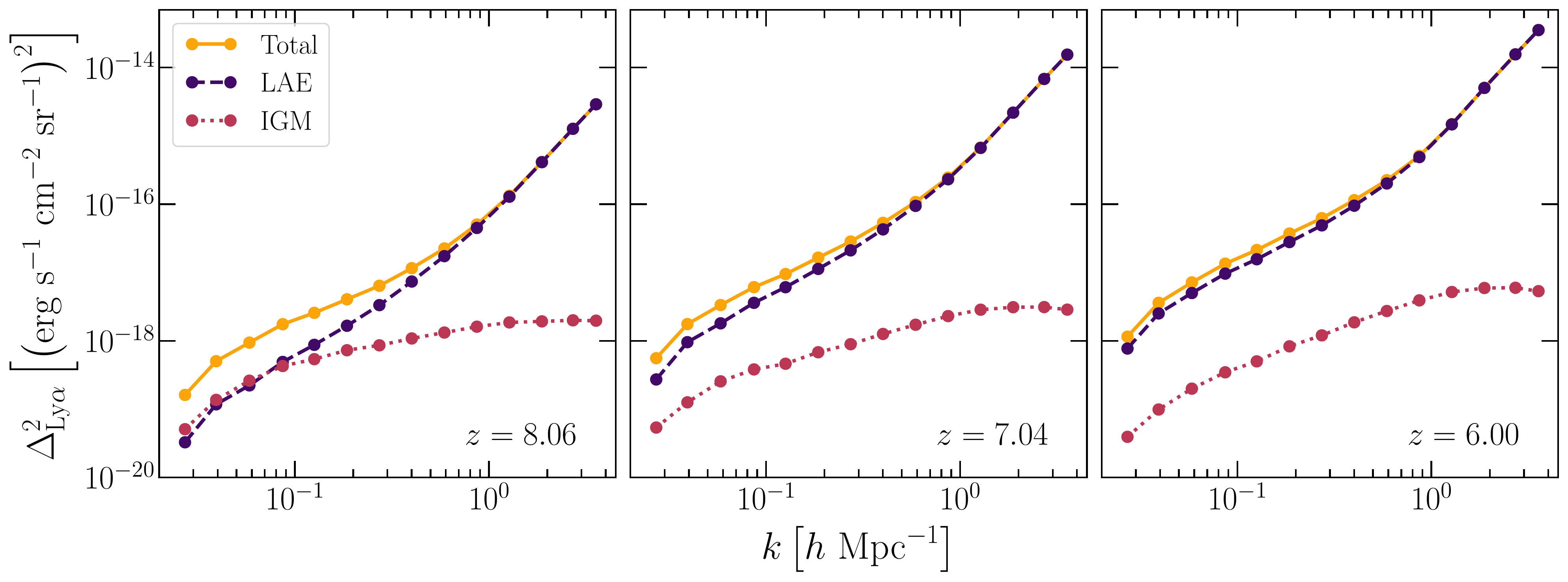}
\caption{\lya \ power spectra across the redshift range of interest. Both the LAE and ionized IGM contributions are included in the calculation. \label{fig:lya_ps}}
\end{figure*}

\subsubsection{Ionized IGM}
\label{subsubsec:ion_IGM}

In this subsection, we describe the \lya\ emission from the ionized IGM. As mentioned at the beginning of this section, we will focus on \lya\ emission in the ionized IGM due to hydrogen recombinations. As with \lya\ emitting galaxies, the ionized bubbles around galaxies also emit \lya\ photons through recombinations of ionized hydrogen. The luminosity density of \lya\ emission in a voxel of the simulation can be defined as
\begin{equation}
    \ell_{\mathrm{rec}}^{\mathrm{IGM}} \left(\textbf{x}, z \right) = n_{\mathrm{rec}} \left(\textbf{x}, z \right) f_{\mathrm{rec}} E_{\mathrm{Ly}\alpha},
\end{equation}
where $E_{\rm Ly\alpha}$ is the rest-frame energy of \lya\ photons, $f_{\rm rec}$ is the fraction of hydrogen recombinations that result in the emission of a \lya\ photon, and $n_{\rm rec}$ is the comoving number density of recombinations occurring in the ionized IGM. The expression for the number density of recombinations is
\begin{equation}
    n_{\mathrm{rec}} \left(\textbf{x}, z \right) = \alpha_{\mathrm{A}} n_{e} \left(\textbf{x}, z \right) n_{\textsc{HII}} \left(\textbf{x}, z \right).
\end{equation}
Here $\alpha_{\mathrm{A}}$ is the case A comoving recombination coefficient, 
\begin{equation}
    \alpha_{\mathrm{A}} \approx 4.2 \times 10^{-13} \left(T_K / 10^{4}\right)^{-0.7} \left(1 + z\right)^3 {\rm cm^3 s^{-1}},
\end{equation}
$n_{e} \left(\textbf{x}, z \right) = x_i n_b$ is the free electron density, and $n_{\textsc{HII}} \left(\textbf{x}, z \right) = n_e A_{\rm He}$ is the comoving number density of ionized hydrogen. With the luminosity density calculated, \cref{eqn:brightness} can be used to calculate the surface brightness of \lya\ in the IGM. The slices through the simulated 21cm and the various components of \lya\ emission cubes are shown in Figure \ref{fig:sims}. The corresponding power spectra are found in Figure \ref{fig:lya_ps}.

In reality, \lya\ emission from the ionized IGM also includes the scattered IGM \lya\ background whose main contributors are X-ray and UV heating, as well as the scattering of Lyman-n photons emitted from galaxies by residual neutral hydrogen in the ionized IGM (\cite{2007MNRAS.376.1680P}). For this work, we chose to neglect the contribution from this \lya\ background as its contribution is subdominant to hydrogen recombination in the ionized IGM (roughly half the mean surface brightness) and the galactic \lya\ contribution (about an order of magnitude lower) \citep{2013ApJ...763..132S, 2017ApJ...848...52H}.

\subsubsection{Ly$\alpha$ Attenuation}
\label{subsubsec:attenuation}

Of course the \lya\ emission does not propagate unimpeded. We must also estimate attenuation of \lya\ by the neutral IGM. The emitted galactic \lya\ experiences an exponential attenuation
\begin{equation}
\label{eqn:lyaatten}
L^{\textrm{gal}}_{\textrm{obs}} = L^{\textrm{gal}} \exp\left[-\tau_{\textrm{Ly}\alpha}\right],
\end{equation}
where $L^{\textrm{gal}}$ is the luminosity of \lya\ emitting galaxies defined in the previous section and $\tau_{\textrm{Ly}\alpha}$ is the optical depth of \lya\ emission at some redshift.

The geography of the ionized bubbles around the emitters determines the extent of the absorption effect.  \lya\ radiation is emitted by some source and as that radiation travels to the edge of the ionized bubble, it is redshifted out of resonance and into the line damping wings, where it has a lower probability of being absorbed by the neutral medium. As reionization progresses and the ionized bubbles around halos grow, the probability of \lya\ radiation being attenuated decreases, as UV photons redshift the further they travel away from the halo before reaching the neutral hydrogen. 

To simulate this behavior, we use a model for the optical depth of \lya\ emission that is defined in \cite{2008MNRAS.385.1348M}. In this model, the optical depth of \lya\ is related to the neutral fraction of IGM that photons encounter along the line-of-sight and the amount by which \lya\ emission is redshifted as it propagates from galactic halos through the neutral IGM. Using our simulation cubes, we estimate a value for the optical depth for \lya\ emission, $\tau_{Ly\alpha}$, by tracing skewers from halos in the simulation through to the edge of the simulation box calculating the redshift on the near, $z_{bi}$, and far, $z_{ei}$, sides of each neutral patch and recording the neutral fraction in that particular voxel, $x_{\textrm{HI}}\left(i\right)$. By tracking these quantities, we calculate the contribution from each neutral hydrogen patch encountered along the line-of-sight of the halo using the approximation \citep{1998ApJ...501...15M}
\begin{align}
\tau_{\textrm{Ly}\alpha} &= \tau_{\textrm{s}} \sum_{i} x_{\textrm{HI}}\left(i\right) \left( \frac{2.02 \times 10^{-8}}{\pi}\right) \left(\frac{1 + z_{bi}}{1 + z_{\textrm{s}}} \right)^{3/2} \nonumber \\
& \hspace{1em} \times \left[ I\left(\frac{1 + z_{bi}}{1 + z_{\textrm{s}}} \right) - I\left(\frac{1 + z_{ei}}{1 + z_{\textrm{s}}}\right) \right].
\end{align}
where $z_{s}$ is the redshift of the \lya\ emitting source. This assumes that the optical line depth at \lya\ line resonance, $\tau_s$, can be approximated as
\begin{equation}
\tau_{\textrm{s}} \approx  6.45 \times 10^5 \left( \frac{\Omega_b h}{0.03}\right) \left( \frac{\Omega_m}{0.3}\right)^{-0.5} \left(  \frac{1 + z_{\textrm{s}}}{10}\right)^{3/2}
\end{equation}
at high-redshifts for a source at some redshift, $z_{s}$, given present-day $\Omega_b$, and $\Omega_m$ (\citealt{1965ApJ...142.1633G, 2001PhR...349..125B}). In the expression above, $I \left(z \right)$ is the helper function,
\begin{equation}
    I\left(x\right) = \frac{x^{4.5}}{1 - x} + \frac{9}{7}x^{3.5} + \frac{9}{5}x^{2.5} + 3 x^{1.5} + 9x^{0.5} - \ln\left(\frac{1 + x^{0.5}}{1 - x^{0.5}}\right),
\end{equation}
derived in \cite{1998ApJ...501...15M}. This expression is only valid for frequencies far from the line center, but is used in this case because the optical depth is so large at the line center that the emission becomes attenuated to the point of being unobservable and therefore gives a fairly accurate approximation for the optical depth at high redshifts.

Once calculated for each halo in the simulation cube, $\tau_{\rm Ly\alpha}$ is applied as an attenuation factor to each halo's intrinsic luminosity through \cref{eqn:lyaatten} before constructing the cubes and calculating the \lya\ power spectrum and 21cm-\lya\ cross-power spectrum for sensitivity calculations in Section \ref{sec:uncertainty}.

\section{Cross-Correlation Statistics} \label{sec:cross-power}
\begin{figure}
\centering
\includegraphics[width=\columnwidth]{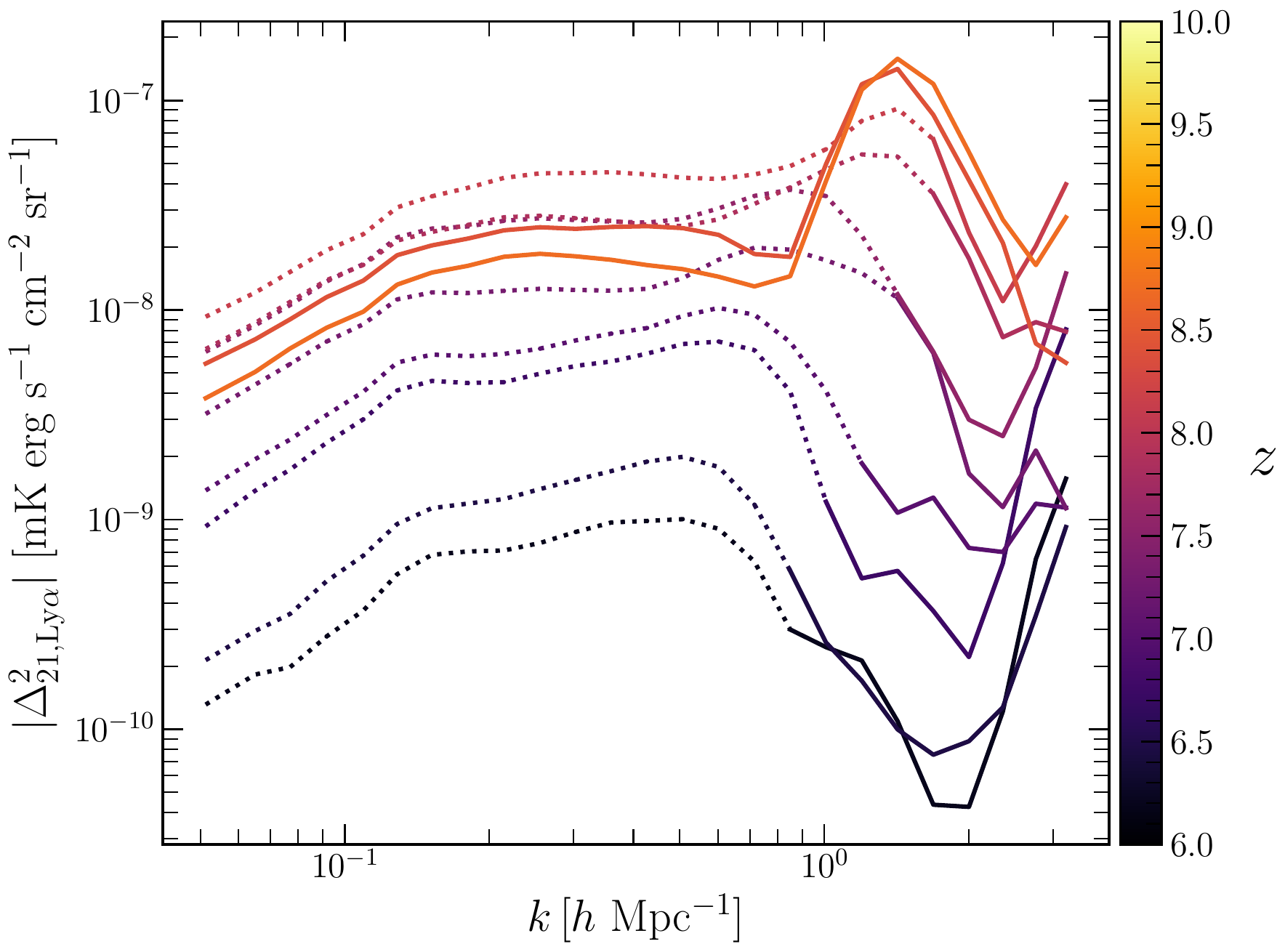}
\caption{The dimensionless 21cm-\lya\ cross-power spectrum. The solid lines on each of the curves represent positive values in the cross-power spectrum, while dotted line on the same curves represent negative values. The cross-power spectrum is expected to turn over from positive to negative on the scale of the mean ionized bubble size at that redshift. We find the cross-power spectrum turns over at increasingly large spatial scales (small $k$-modes) as reionization progresses, tracing the growth of ionized bubbles. \label{fig:cps_plot}}
\end{figure}

As direct observation of 21cm image cubes will require next generation sensitivity and precision, current experiments are opting to focus on the power spectrum, a measurement of statistical fluctuations on various spatial scales. The power spectrum, $P \left({\bf k} \right)$, is formally defined as
\begin{equation}
\langle \widetilde{\delta T}_{b} ({\bf k}) \widetilde{\nu I}_{\mathrm{Ly}\alpha} ({\bf k'}) \rangle \equiv (2 \pi) ^3 \delta^D \left( {\bf k} - {\bf k'} \right) P_{21, \mathrm{Ly}\alpha} \left({\bf k} \right),
\end{equation}
where $\widetilde{\delta}$ is the Fourier transform of some fluctuation field, in this case either 21cm or \lya, $\delta^D$ is the Dirac delta function. More commonly used in the literature is the dimensionless power spectrum,
\begin{equation}
    \Delta^2_{21, \mathrm{Ly} \alpha} \left( k \right) = \frac{k^3}{2 \pi ^2} P_{21, \mathrm{Ly} \alpha} \left( k \right),
\end{equation}
which represents the contribution to the co-variance in bins of $k$, and in which the $k^3$ cancels the spatial units from the power spectrum. Figure \ref{fig:cps_plot} shows the cross-power spectrum for a range of redshifts from $z \approx 6-8$.

In the power spectrum it is difficult to decouple the total power of the two fields from their correlation. Though we can't directly measure the correlation it is useful to inspect it in simulation as a way to understand the dependence of the correlation on size scale. We expect the correlation coefficient

\begin{equation}
  r_{21, \rm Ly\alpha} \left( k \right) = \frac{P_{21, \rm Ly \alpha} \left( k \right)}{\sqrt{P_{21} \left( k \right) P_{\rm Ly \alpha} \left( k \right)}},
\end{equation}

to have an anti-correlation $(r < 0)$ on large scales dominated by bubbles which approaches zero at a scale roughly corresponding to the average size of ionized regions.

\begin{figure}
\centering
\includegraphics[width=\columnwidth]{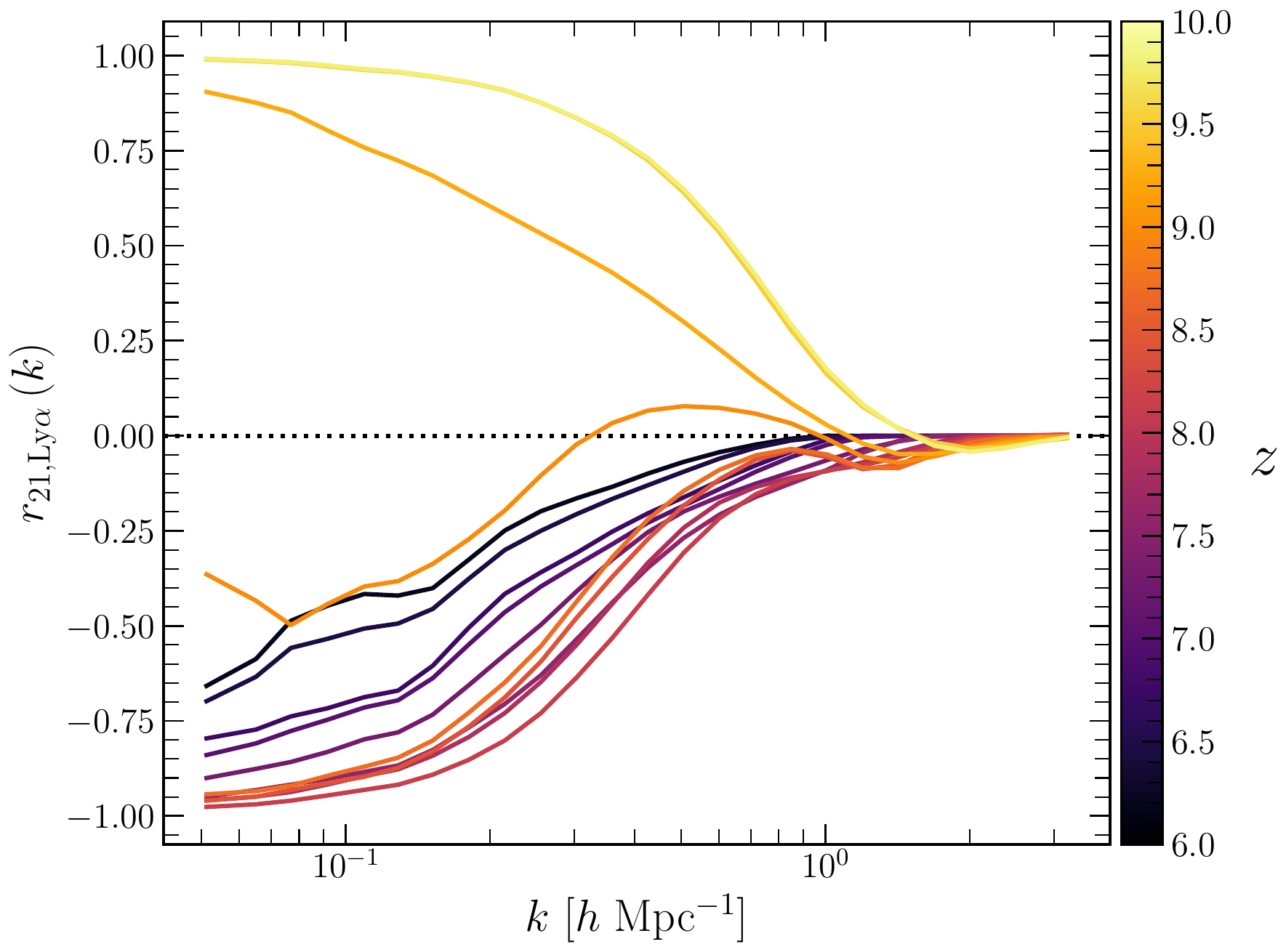}
\caption{The cross-correlation coefficient as a function of spatial scale. Here, a cross-correlation coefficient value of $-1$ indicates that 21cm and \lya\ emission are completely anti-correlated, while a value of 0 indicates no correlation. At a fixed spatial scale, $r\left(k\right)$ becomes less negative between $z\approx8$ and $z\approx6$ as reionization progresses, which tracks the growth of ionized bubbles around galaxies. Above redshift of $z \sim 8$ we see an increase in the correlation at large scales which is explained by few galaxies forming in overdense regions that haven't yet been ionized leading to a correlation between 21cm and \lya \ fields\citep{2009ApJ...690..252L}.
 \label{fig:ccc_plot}}
\end{figure}

The correlation (shown in Figure \ref{fig:ccc_plot}) behaves as expected, transitioning from uncorrelated on small-scales to anti-correlated on large-scales, where the fluctuations from the two fields are not overlapping. It is also interesting to note that the cross-correlation coefficient progresses from generally anti-correlated at high-redshifts to generally uncorrelated towards the end of reionization. This change traces the growth of ionized bubbles through the process of reionization. The scales at which the cross-correlation coefficient transitions from uncorrelated to anti-correlated represents the size of typical ionized bubbles around galaxies, which matches results seen in previous literature \citep{2009ApJ...690..252L, 2017ApJ...848...52H, 2018MNRAS.479.2754K}.

\section{Detectability of the Cross-Power Spectrum} \label{sec:uncertainty}

    \subsection{Thermal Noise Contribution} \label{subsec:thermal}
The feasibility of a 21cm-\lya\ cross-power spectrum measurement is governed by sources of uncertainty associated with both measurements independently and their overlap. Calculation of thermal error bars captures both. In this work, we use HERA as our 21cm instrument, and SPHEREx and CDIM individually for \lya . 

The noise on one particular $k$-mode in the cross-power spectrum depends on noise from both the \lya\ and 21cm measurements, and therefore the uncertainty from both measurements must be calculated. The variance on the cross-power spectrum due to contributions from both instruments and sample variance from the cross power spectrum is,
\begin{align}
  \sigma^2&_{21, \rm Ly\alpha} \left( \textbf{k} \right) = \frac{1}{2} \left[ P^2_{21, \rm Ly\alpha}\left( \textbf{k} \right) + \sigma_{21}\left( \textbf{k} \right) \sigma_{\rm Ly \alpha}\left( \textbf{k} \right) \right] \nonumber \\
  &\propto P^2_{21, \rm Ly\alpha} + \left(P_{21} + P_{21, \mathrm{N}}\right) \left(P_{\rm Ly \alpha} + P_{\rm Ly \alpha, \mathrm{N}} \right).
  \label{eqn:sensitivty}
\end{align}
Here, $P_{21, \rm Ly\alpha}$ is sample variance due to the cross-power spectrum, $P_{21}$ is sample variance from the 21cm signal, $P_{\rm Ly\alpha}$ is sample variance from the $\rm Ly\alpha$ measurement, and $P_{21, N}$ and $P_{{\rm Ly\alpha}, N}$ are thermal noise uncertainty terms from 21cm and $Ly\alpha$ measurements respectively.

To calculate the thermal noise for the 21cm observation, we use the method described by \cite{2013AJ....145...65P}\footnote{We used the updated version of the code at \url{https://github.com/steven-murray/21cmSense}. This version adds support for modern configuration formats like YAML, significant unit-testing, documentation and tutorials, and greatly modularises the code.}. In this method, \textit{uv}-coverage of the observation is taken into account using the exact layout of HERA and applying Earth-rotation synthesis to simulate changing \textit{uv}-bins sampled by each pair of antennas. This \textit{uv}-coverage then dictates the exact $k_{\perp}$ resolution of the instrument while its spectral resolution sets the $k_{\parallel}$ resolution. We then perform a spherical average over the $k_{\parallel}$ and $k_{\perp}$ bins to identify the observation time, $t_{\rm int}$, associated with each $k$-bin. As described in \cite{2014ApJ...782...66P}, the power spectrum error depends on the observation time $t_\textrm{int}$ and the system temperature $T_{\rm sys}$ according to
\begin{equation}
P_{21, \textrm{N}} = X^2 Y \frac{T^2_{\textrm{sys}}}{2 t_{\mathrm{int}}} \frac{\Omega_{\textrm{p}}^2}{\Omega_{\textrm{pp}}}.
\end{equation}
Here, $X^2Y$ converts bandwidth and solid angle to their relevant cosmological scale equivalents. The field of view also factors in via $\Omega_p$, the solid angle of the primary beam, and $\Omega_{\rm pp}$, the solid angle of the primary beam squared. For the  thermal noise estimates above, the fiducial observing parameters for HERA ($t_{\rm int} = 1000 \ \rm hr$, $T_{\rm sys} = 100 + 120 \left(\nu / 150\right)^{-2.55}$, and $B = 8 \ \rm MHz$) are assumed \citep{2017PASP..129d5001D}.

The thermal noise associated with infrared intensity mapping experiments (SPHEREx and CDIM in this work) can be written as
\begin{equation}
P_{\mathrm{Ly}\alpha, \mathrm{N}} = \left(\nu_{\rm obs} \sigma_{\rm N}\right)^2 V_{\mathrm{vox}} \textrm{W}_{\rm Ly\alpha},
\end{equation}
where $\nu_{\rm obs}$ is the observed frequency of \lya, $V_{\mathrm{vox}}$ is the comoving voxel volume and $\textrm{W}_{\rm Ly\alpha}$ is the window function defined in \cite{2011ApJ...741...70L}, which accounts for limitations in the spectral and spatial instrumental resolution of the instrument. In the equation above, we take $\sigma_{\rm N} = 10^{-18} \ {\rm erg \ s^{-1} \ cm^{-2} \ sr^{-1} \ Hz}$, which is consistent with the 5$\sigma$ value reported by \cite{2016arXiv160607039D} for SPHEREx, and $\sigma_{\rm N} = 1.5 \times 10^{-19} \ {\rm erg \ s^{-1} \ cm^{-2} \ sr^{-1} \ Hz}$ for CDIM \cite{2019BAAS...51g..23C}.  In addition to these values, we also explore more optimistic sensitivity values assuming deeper integrations than the minimum system requirements for both SPHEREx and CDIM as an exploratory measure for the requirements of a detection of the cross-power spectrum. For these optimistic values, we take $\sigma_{\rm N} = 3 \times 10^{-20} \ {\rm erg \ s^{-1} \ cm^{-2} \ sr^{-1} \ Hz^{-1}}$\footnote{\url{https://github.com/SPHEREx/Public-products/}} for SPHEREx, and $\sigma_{\rm N} = 1.5 \times 10^{-21} \ {\rm erg \ s^{-1} \ cm^{-2} \ sr^{-1} \ Hz^{-1}}$ for CDIM \cite{2017ApJ...848...52H}. We do expect the nominal values to be the more realistic values measured, but better thermal noise values are possible given certain instrument assumptions \citep{2018arXiv180505489D, 2021ApJS..252...24S}.

\cref{eqn:sensitivty} is defined for the unaveraged power spectra. In practice, we are more interested in the spherically averaged noise power spectrum. To obtain the variance on the spherically averaged cross-power spectra, we use
\begin{equation}
    \frac{1}{\sigma^2 \left(k \right)} = \sum_{\textbf{k} \in k} \frac{N_m}{\sigma^2 \left( \textbf{k} \right)},
\end{equation}
where $N_m$ is the number of modes within a particular $k$-bin, which are explicitly counted when averaging. The results of these cuts can be found in Figure \ref{fig:error_budget}. A discussion of the results of this figure coupled with the different foregrounds strategies in the following section.

    \subsection{Foreground Contamination}\label{subsec:foregrounds}
In addition to the thermal noise and limited spectral and spatial resolution of each instrument, we also like to explore the effect of foregrounds on our ability to effectively measure the cross-power spectrum. Cross-correlation of reionization-era 21cm observations with \lya\ intensity mapping surveys have the advantage that 21cm foregrounds are expected to have no correlation with low redshift interloper lines that affect high-redshift \lya\ measurements. Because the two are uncorrelated, power from the each of these foregrounds should not be directly added to the cross-power spectrum. However, while bright foregrounds are not expected to contribute to the amplitude of the cross-power spectrum, they do contribute to the total variance on the cross-power spectrum if not removed. To truly be confident in a detection of the cross-power spectrum, foregrounds must be accounted for.

While \lya\ intensity mapping experiments do have to contend with interloper lines at lower redshifts, such as H$\alpha$, O\textsc{ii}, and O\textsc{iii} that are orders of magnitude brighter than the infrared background, much work has gone into effectively removing these lines. The technique for removing foreground sources simply involves applying a cutoff flux and removing all pixels whose amplitude falls above that threshold. Previous work has shown that for a SPHEREx-like infrared satellite with spectral resolution $R\approx 41.5$ only 3\% of pixels would need to be removed to bypass all foreground interlopers \citep{2014ApJ...786..111P, 2014ApJ...785...72G}. Similarly, \cite{2017ApJ...846...21F} found that for a CDIM-like experiment $R\approx 300$ only 0.1\% of pixels would need to be removed to significantly lower the amplitude of the interloper to orders of magnitude below the \lya\ power spectrum. For this reason, in this work we ignore the effect infrared foreground removal would have on decreasing the amplitude of the cross-power spectrum and instead focus on the effect of 21cm foreground removal on the cross-power spectrum.

\begin{figure*}
\centering
\includegraphics[width=.95\textwidth]{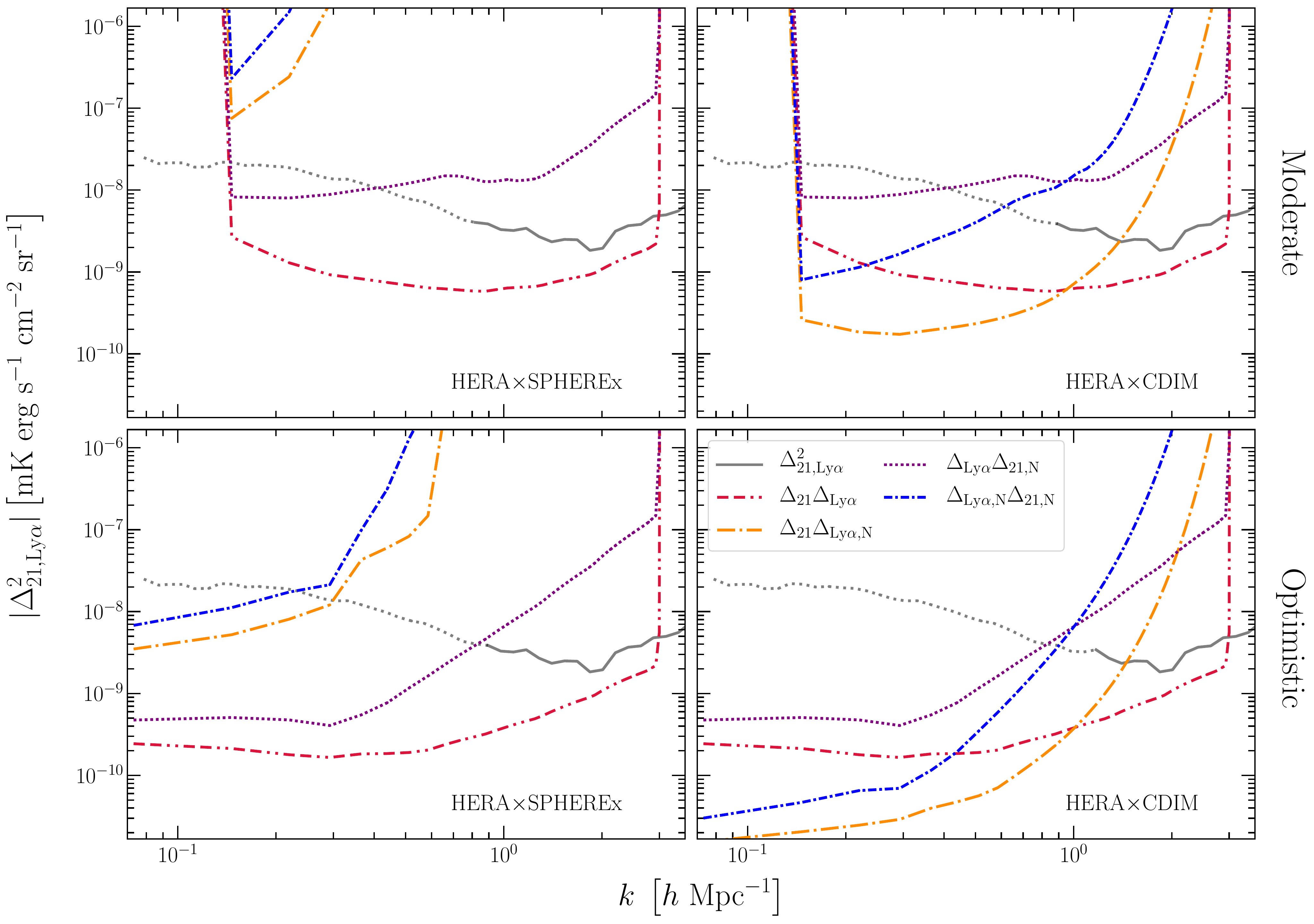}
\caption{Error budget of the sensitivity of the cross-power spectrum at $z = 7.04$ between the HERA 21cm experiment and Ly$\alpha$ observations by SPHEREx and CDIM for the deep integration case. Here, the cross terms defined in Eq. \ref{eqn:sensitivty} are plotted individually to isolate the sources of error on a measurement of cross-power spectrum. The relevant sensitivity threshold is the quadradic sum of all terms which is dominated by the whichever term is largest. The simulated cross-power spectrum is shown in grey with negative values in dashes.  Moderate 21cm foreground filtering in shown at top and optimistic in the bottom row. See section \ref{subsec:foregrounds} for a complete discussion. \label{fig:error_budget}}
\end{figure*}
The foreground wedge is a well-documented feature in 21cm literature \citep[eg.][]{2010ApJ...724..526D, 2012ApJ...752..137M, 2014PhRvD..90b3018L} and filtering matched to this shape has been identified as a potential method of removing bright 21cm foregrounds. The foreground wedge is a feature that appears in the cylindrically-averaged 21cm power spectrum, arising as a product of spectrally smooth foregrounds and the chromatic response of the interferometer. Because of their smooth spectral structure, bright foregrounds are confined to low-order Fourier modes, thus constraining their power to low $k_{\parallel}$ values. Experiments such as HERA have leveraged the fact that the edge of the foreground wedge is dependent on the baseline length of two antennas by building densely packed arrays that sample lower $k_{\perp}$ values, thus increasing the EoR window. The approximate relationship defining the $k_{\parallel}$-edge of the wedge can be written as
\begin{equation}
  k_{\parallel, \mathrm{max}} = \frac{D_M}{D_H} \frac{E \left( z \right)}{\left(1 + z \right)} \sin{\theta_0} k_{\perp}
\end{equation}
where the characteristic angle, $\theta_{0}$, is the instrument field of view, $k_{\perp}$ is the Fourier mode dependent on the distance between two dishes, and $k_{\parallel, \mathrm{max}}$ corresponds to the maximum $k_{\parallel}$ value dominated by bright foregrounds.

Typically, the safest assumption to make is that $\theta_{0} = \pi / 2$, which corresponds to a field of view that includes bright foregrounds at the horizon, far from the pointing center. In practice, the wedge can extend even beyond the horizon given imperfectly calibrated chromaticity and internal instrument systematics, such as cable reflections and cross-coupling effects \citep{2013ApJ...768L..36P, 2020ApJ...888...70K}. However, work by \citep{2014ApJ...782...66P} has argued that outside the field of view, foreground contamination is sufficiently attenuated by the primary beam so as not to corrupt the cosmological 21cm signal. If this is achievable in practice, it would increase the size of the EoR window and provide a significant sensitivity boost to a cross-power spectrum measurement.

To investigate the effect of the foreground wedge on the ability to measure the cross-power spectrum, we adopt two treatments of 21cm foregrounds described in \cite{2014ApJ...782...66P}: a moderate foreground treatment, where the foreground wedge extends to the horizon with a horizon buffer added to account for improper calibration, and an optimistic foreground treatment, where the wedge is confined to the first null in the primary beam of the instrument. In both of these treatments, we remove all $\mathbf{k}$-modes that fall within the foreground wedge of the cylindrically-averaged power spectra before averaging down to the spherically-averaged power spectra. We compute each of the cross terms defined in Eq, \ref{eqn:sensitivty} shown in Figure \ref{fig:error_budget}. There are two signals for four cross terms. 
The total error is the quadrature sum of these terms which is well approximated by the largest error on the plot. For SPHEREx the two largest components are the $\rm Ly\alpha, N$ noise-noise term  followed closely by the $\rm 21cm-Ly\alpha$ cross term. This suggests the sensitivity of the Ly$\alpha$ measurement is the limiting factor.  For CDIM, however the reverse is true. At the most sensitive $k$-modes the dominant source of error is the correlation between $\rm Ly\alpha$ sample variance and 21cm thermal noise uncertainty, suggesting that with CDIM's greater range of available modes the 21cm sensitivity becomes the limiting factor.

The other strong factor at work is the 21cm foreground treatment; filtering modes measured by both instruments reduces sensitivity with a strong dependence on $k$.
In the moderate treatment there is a sharp rise in the uncertainty towards lower $k$s. This is due to the complete loss of large scales in the 21cm wedge filter.   We can also see that sensitivity improves across the board when the filter is relaxed reflecting the fact that a range of $k_\perp$ are being included.  For SPHEREx, the filter has the largest impact on the cross term between the 21cm signal and $\rm Ly\alpha$ noise. With the limited number of modes overlapping between HERA and SPHEREx, the loss of modes to the foreground filter is keenly felt.  With its much wider range of available modes wedge filtering has a much smaller impact on the correlation with CDIM.

\subsection{Sensitivity Estimates}
\begin{figure}
\centering
\includegraphics[width=\columnwidth]{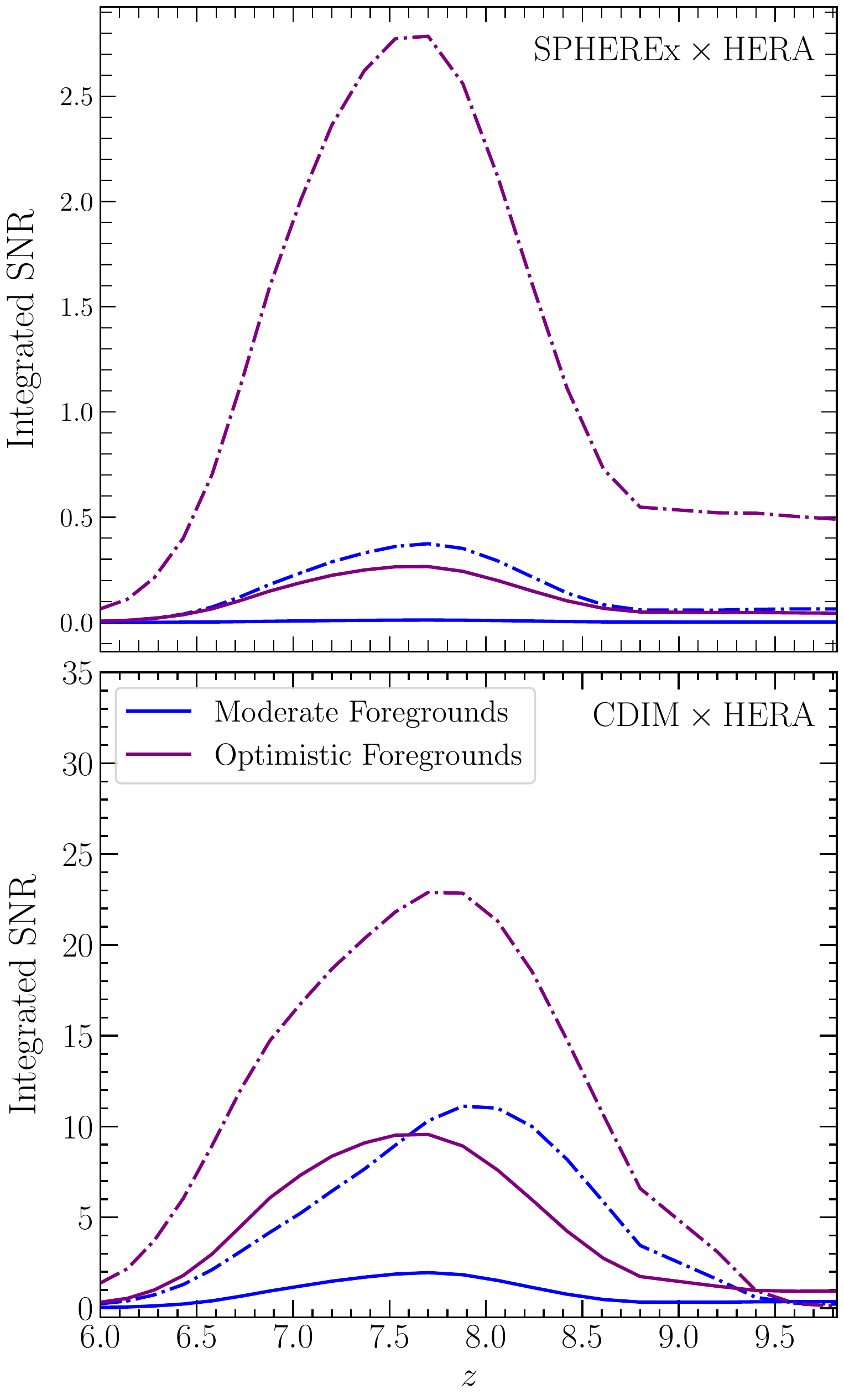}
\caption{Integrated signal-to-noise on the cross-power spectrum as a function of redshift. For each panel, the blue line represents SNR estimates on the cross-power spectrum in the moderate foreground case, while the purple line represents the optimistic foreground case. In addition to different foreground strategies, we also differentiate between the minimum thermal noise requirement for each instrument (solid lines) and a more optimal deep integration (dotted-dashed lines) for each \lya \ experiment\label{fig:snr}}
\end{figure}

With the thermal noise, spectral and spatial resolution effects, and foregrounds taken into account, we can examine the sensitivity these instrument pairs have to the cross-power spectrum. Using the cross-power spectrum and the noise cross-power spectrum calculated previously, we can calculate the total signal-to-noise ratio across the cross-power spectrum by summing across the $k$-bins using the expression
\begin{equation}
    \mathrm{SNR^2_{total}} = \sum_i \mathrm{SNR}^2_i = \sum_i \left(\frac{P_{21, \mathrm{Ly}\alpha} \left(k_i \right)}{\sigma_{21, \mathrm{Ly}\alpha} \left(k_i \right)}\right)^2,
\end{equation}
where the index, $i$, iterates through each of the $k$-bins. This signal-to-noise calculation was done for each redshift bin, for both instruments, and both foreground treatments. These calculated signal-to-noise ratios for HERA/SPHEREx and HERA/CDIM cross-power spectra are found in Figure \ref{fig:snr}.

The overall SNR prediction (Figure \ref{fig:snr}) tells us that cross-correlation with SPHEREx requires an optimistic treatment of the 21cm foregrounds and deeper integrations for SPHEREx than the minimum requirements to make a detection of the cross-power spectrum; which here means making noise-limited measurements at delay bins up to the first null of the beam. SPHEREx and HERA sample different $\mathbf{k}$-mode ranges which overlap best at small $k$. Meanwhile, the correlation power spectrum, like the auto spectrum, is roughly flat in $k$. Looking further into the future, a HERA/CDIM-like cross-correlation may be possible even if the entire wedge is excluded from the 21cm data. This is made possible by the much larger sensitivity of CDIM at 21cm $k$-modes which are foreground free. 

\section{Summary} \label{sec:summary}
 We have tested the feasibility of detecting large scale structure during reionization by cross-correlating HERA with two future infrared intensity mapping satellites, SPHEREx and CDIM. In the near future, using cross-correlations between HERA and SPHEREx, we find that the cross-power spectrum may be detectable from $z \approx 7-8.5$ in the fiducial 21cm model, but only with aggressive removal of 21cm foregrounds and deeper thermal noise integrations than the minimum system requirements. This is due to a lack of overlapping sensitive modes between the two instruments that is only remedied by aggressive foreground removal or deep integrations from either instrument. A HERAxSPHEREx cross-correlation will likely set upper limits on the intensity of the cross-power spectrum, as well as some constraints on astrophysical parameters, but a detection of the cross-power spectrum will be challenging without better control of systematics and improved foreground subtraction. More forward-looking, we also show that a HERAxCDIM  cross-power spectrum measurement ought to have sufficient signal-to-noise to detect the fiducial model across a significant portion of reionization ($z \approx 6 - 9; x_{\textrm{HI}} \approx 0.01 - 0.75$), assuming a fairly aggressive foreground treatment and slightly deeper integrations than the minimum system requirements.
 
 More work is needed to extend this initial study.  We treated foregrounds by removing likely contaminated modes. One could imagine that small residual foregrounds could cancel in cross-correlation, thus opening up more modes. While it is likely that residual foregrounds from even the best levels of foreground modeling and subtraction currently in use would dominate the error budget of the cross-power spectrum, the relatively low noise contribution from CDIM may eventually allow the cross-power spectrum to be detectable even in cases of imperfect 21cm foreground removal at large scales.

Potentially the most exciting aspect of cross-correlating 21cm and \lya\  measurements will be determining its ability to constrain astrophysical and cosmological model parameters using the cross-power spectrum. Synergies between line intensity mapping experiments will help drive constraints on parameter estimates by breaking down expected degeneracies. We leave a study of astrophysical and cosmological parameter estimation for future work.

\section*{Acknowledgements}
The authors would like to thank Alexander van Engelen and Judd Bowman for their helpful comments on an early draft. This material is based upon work supported by the National Science Foundation under Grant \#1636646 and \#1836019 and institutional support from the HERA collaboration partners. This research is funded in part by the Gordon and Betty Moore Foundation. HERA is hosted by the South African Radio Astronomy Observatory, which is a facility of the National Research Foundation, an agency of the Department of Science and Innovation.

\subsection*{Software}
This work was enabled by a number of software packages including \pkg{21cmSense} \citep{2013AJ....145...65P}, \pkg{astropy} \citep{2013A&A...558A..33A}, \pkg{matplotlib} \citep{Hunter:2007}, \pkg{numpy} \citep{oliphant2006guide}, \pkg{powerbox} \citep{2018JOSS....3..850M}, and \pkg{scipy} \citep{2020SciPy-NMeth}.

\subsection*{Data Availability}
The 21cm and \lya \ simulation cubes and noise curves used in this study can be provided upon reasonable request to the main author.

\bibliography{references, software}
\bibliographystyle{aasjournal}
\end{document}